\documentclass[10pt]{article}
\usepackage{psfig,epsf}
\usepackage{a4wide,graphicx}
\usepackage{cite}

\def\J{J/\Psi}
\def\MeV{\textrm{ MeV}}

\def\J{J/\Psi}
\newcommand{\be}{\begin{equation}}
\newcommand{\ee}{\end{equation}}
\newcommand{\ba}{\begin{eqnarray}}
\newcommand{\ea}{\end{eqnarray}}

\begin{document}

\markboth{L.~Roca, J.~E.~Palomar, E.~Oset, H.~C.~Chiang}
{Unitary chiral dynamics in $J/\Psi$ decays into $VPP$ 
and the role of the scalar mesons}

%
%

\title{Unitary chiral dynamics in $J/\Psi$ decays into $VPP$ 
and the role of the scalar mesons}

\author{
L.~Roca$^1$, J.~E.~Palomar$^1$, E.~Oset$^1$ and H.~C.~Chiang$^2$\\
{\it {\small $^1$Departamento de F\'{\i}sica Te\'orica and IFIC,
Centro Mixto Universidad de Valencia-CSIC,}} \\ 
{\it {\small Institutos de
Investigaci\'on de Paterna, Aptdo. 22085, 46071 Valencia, Spain}}\\ 
{\it {\small $^2$Institute of High Energy Physics, Chinese Academy of Sciences, Beijing
100039, China}}
}

\maketitle


\begin{abstract}
We make a theoretical study of  the $\J$ decays into
$\omega\pi\pi$, $\phi\pi\pi$, $\omega K \bar{K}$ and $\phi
K\bar{K}$ using the techniques of the chiral unitary approach
stressing the important role of the scalar resonances
dynamically generated through the final state interaction of the
two pseudoscalar mesons. We also discuss the importance of new
mechanisms with intermediate exchange of vector and axial-vector
mesons and the role played by the OZI rule in the $\J\phi\pi\pi$
 vertex, quantifying its effects. The results nicely
reproduce the experimental data for the invariant mass
distributions in all the channels considered. 

\end{abstract}

\vspace{1cm}

The $\J$ decay into a pseudoscalar meson pair and a vector meson
has been claimed to be one of the most suited reactions to study
the long controversial nature of the scalar mesons, on which the
interpretation as $q\bar{q}$ mesons or as meson-meson molecules
has mainly centered the discussion.  In the last years, a chiral
unitary coupled channel approach  
\cite{Oller:2000ma} has proved to be
successful in describing meson-meson interactions in all
channels up to energies $\sim1.2\textrm{ GeV}$, far beyond the
natural limit  of applicability of the standard Chiral
Perturbation Theory (ChPT), which is $\sim500\MeV$ where the
pole of the lightest resonance, the $\sigma$ meson, appears. In
this approach the scalar mesons 
rise up naturally as dynamically generated
resonances, in the sense that, without being included as
explicit degrees of freedom, they appear as poles in the s-wave
meson-meson scattering amplitudes. The aim of the present work
\footnote{All the details of the model can be found in \cite{yo},
on which the present contribution is based}
is to make a consistent and comprehensive description of the
$\J\to VPP$ decays, including all the mechanisms able to
influence the region of pseudoscalar pair invariant masses up to
$\sim 1.1\textrm{ GeV}$, addressing the role played by the
scalar mesons and the OZI rule. 
 Following the framework
of the chiral unitary approach, we will implement the final
meson-meson state interaction in order to generate dynamically
the scalar resonances involved.

\begin{figure}
\centerline{\protect\hbox{
\psfig{file=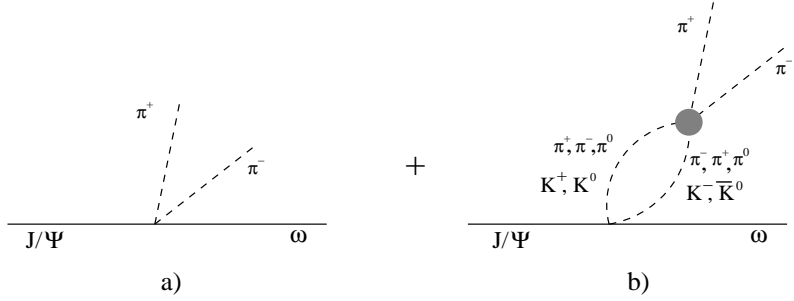,width=0.7\textwidth,silent=}}}
\caption{Diagrams with direct $\J VPP$ vertex at tree level, a), and with
the iterated meson loops, b). }
\label{fig:chiral_loops}
\end{figure}

The first mechanisms considered are those involving a
direct coupling of the $\J$ to the two pseudoscalars and the
vector, implementing the final state interaction of the
pseudoscalars pair, as  is depicted in
Fig.~\ref{fig:chiral_loops} for the $\J\to\omega\pi^+\pi^-$
channel. 
The thick dot in Fig.~\ref{fig:chiral_loops} means that one is
considering the full $\pi\pi(K\bar{K})\to \pi^+\pi^-$
$t$-matrix, involving the loop resummation of the Bethe-Salpeter
equation and no just the lowest order
$\pi\pi(K\bar{K})\to \pi^+\pi^-$ amplitude. Actually this loop
resummation is what dynamically generates the scalar resonances. 
The $\J VPP$ vertex can be constructed using $SU(3)$ arguments to relate
the different isospin channels and parameterizing the amplitudes
in a way which clearly manifest the role played by the OZI rule
in this direct vertex, accounted for
by means of a parameter $\lambda_\phi$ such that, if the OZI rule
would be exact, then $\lambda_\phi$ would be zero.

\begin{figure}[tbp]
\centerline{\hbox{\psfig{file=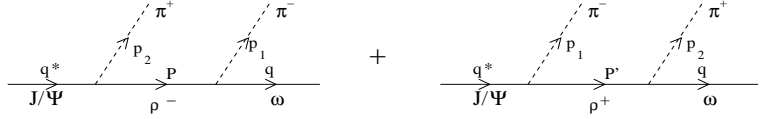,width=10cm}}}
\caption{\rm 
Diagrams for the tree level mechanisms with sequential vector meson
exchange.}
 \label{fig:VMD_tree_w}
\end{figure}

In analogy with the approach in Ref.~\cite{rocaphi} (and in some
sense to \cite{rocaeta}), we consider next
the mechanisms involving the
sequential exchange of a vector or axial-vector meson as depicted
in  Fig.~\ref{fig:VMD_tree_w} for the vector meson case for the
$\omega\pi\pi$ channel.
The vertices needed are obtained from suitable phenomenological
Lagrangians (see {\cite{yo,rocaphi,axials}).
In
 Figs.~\ref{fig:VMD_loop_pions_w} and ~\ref{fig:VMD_loop_kaons_w}
we show the mechanisms with the implementation of the FSI of the
pseudoscalar mesons.
\begin{figure}[tbp]
\centerline{\hbox{\psfig{file=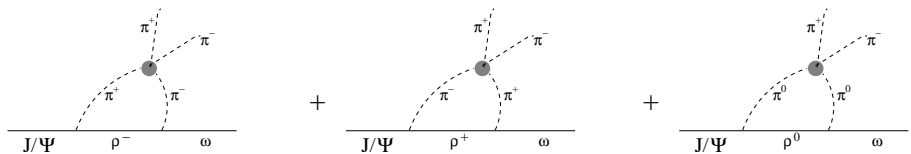,width=12cm}}}
\caption{\rm 
Sequential vector meson exchange diagrams with final state
interaction of pions}
\label{fig:VMD_loop_pions_w}
\end{figure}
\begin{figure}[tbp]
\centerline{\hbox{\psfig{file=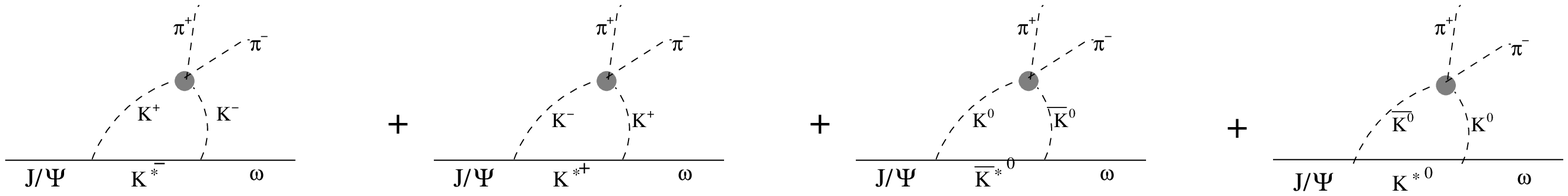,width=12cm}}}
\caption{\rm 
Sequential vector meson exchange diagrams with final state
interaction of kaons}
\label{fig:VMD_loop_kaons_w}
\end{figure}

The only two free parameters in our model are the coupling of
the direct $\J VPP$ vertex and the OZI rule violation parameter,
$\lambda_\phi$. Fitting our model to the $\J\to\omega\pi^+\pi^-$
and $\J\to\phi\pi^+\pi^-$ experimental data we obtain values of
$\lambda_\phi$ clearly different from zero ($\lambda_\phi=0.12,
0.20$) and reasonably
smaller than one, what manifests the OZI rule violation within
reasonable values.
In Fig.~5, left column,
 we show the results for $\omega\pi\pi$ and
$\phi\pi\pi$ channels
including an estimation of the theoretical error band.
\begin{figure}[tbp]
\centerline{\hbox{\psfig{file=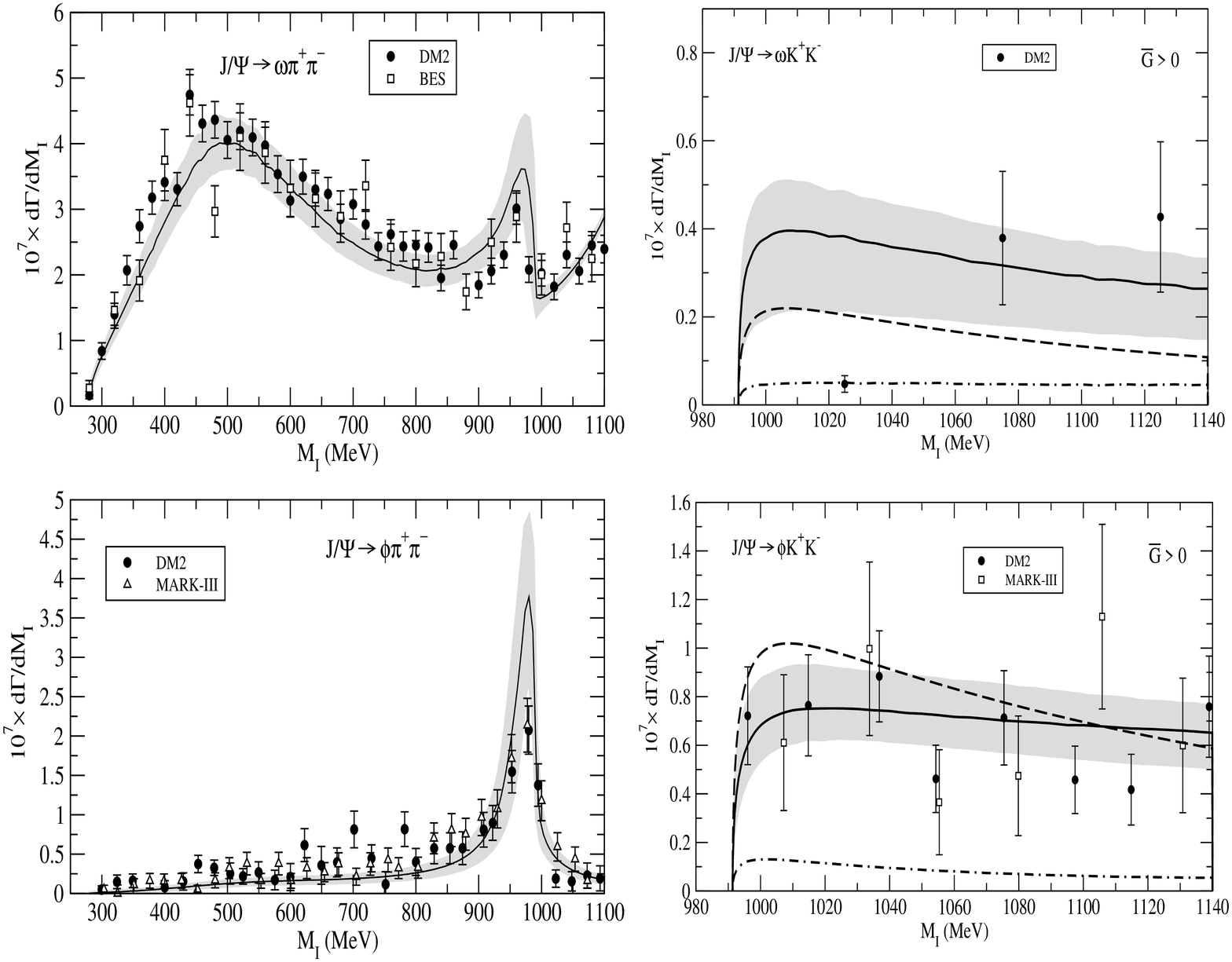,width=12cm}}}
\caption{}
\end{figure}
The experimental data shown in the figures have been obtained
from \cite{WuBES,Augustin,Falvard,Lockman}.
 Concerning the scalar mesons, it is important to stress first
that the final shape and strength of the bump appearing in the
$\pi\pi$ invariant mass distribution in the $\omega\pi\pi$
channel at $\sim 500\MeV$ is determined by a subtle interference
between the final state interaction  in the direct $\J VPP$
mechanisms and the tree level direct $\J VPP$ decay. This means
that the shape and position of the bump does not directly
represent the physical properties of the $\sigma$ meson, since
it is distorted due to interferences with other terms not
related to the $\sigma$ meson.
 Therefore one has to be
extremely careful when using the experimental
 data to extract the physical
$\sigma$ meson properties by fitting Breit-Wigner-like shapes.
On the other hand, the relative weights of the $f_0(980)$  and
the $\sigma$ meson are well reproduced in both the 
$\omega\pi\pi$ and $\phi\pi\pi$ channels in spite of their large
difference in these channels. This relative weight is mainly
determined by the OZI rule violation parameter and the
interferences of the "direct" terms with the other mechanisms,
specially in the $f_0(980)$ region. In our model, since the
scalar mesons are dynamically generated through the resummed
meson-meson amplitude, the relative weight between the
$f_0(980)$ and the $\sigma$ mesons is related to the relative
weight between the $K\bar K\to\pi\pi$ and $\pi\pi\to\pi\pi$, in
$I=0$, scattering amplitudes. Specially remarkable is the fair
agreement in the $f_0(980)$ region of the $\omega\pi\pi$ channel
despite the smallness of the bump and the fact that many
mechanisms contribute in this region.

Finally, in Figs.5, right column, we apply
 our results to the  $\omega K\bar K$
and $\phi K\bar K$ decay channels, obtaining a fair agreement
without introducing any extra freedom in the model. This is a
nice test of the present model, both reproducing the absolute
strength, and also the shape, which shows much strength close to
threshold as a reflection of the proximity of the $f_0(980)$
resonance below threshold.

In conclusion, we have obtained a good description of these
interesting $\J$ decays combining phenomenological Lagrangians
and the techniques of the chiral unitary approach to implement
the final state rescattering of the pseudoscalar pairs,
quantifying the controversial non-trivial role of the scalar
mesons and the violation of the OZI rule.
The fact that once more one is able to reproduce the shape and
strength of the $f_0(980)$ and the $\sigma$ resonances without the
need to introduce them as explicit degrees of freedom provides an
extra support to the idea of the nature of these resonances as
dynamically generated from the interaction of the mesons.


\begin{thebibliography}{0}

\bibitem{Oller:2000ma}
J.~A.~Oller, E.~Oset and A.~Ramos,
Prog.\ Part.\ Nucl.\ Phys.\  {\bf 45} (2000) 157.

\bibitem{yo}
L.~Roca, J.~E.~Palomar, E.~Oset and H.~C.~Chiang,
arXiv:hep-ph/0405228.

\bibitem{rocaphi}
J.~E.~Palomar, L.~Roca, E.~Oset and M.~J.~Vicente Vacas,
Nucl.\ Phys.\ A {\bf 729} (2003) 743
[arXiv:hep-ph/0306249].

\bibitem{rocaeta}
E.~Oset, J.~R.~Pelaez and L.~Roca,
Phys.\ Rev.\ D {\bf 67} (2003) 073013
[arXiv:hep-ph/0210282].

\bibitem{axials}
L.~Roca, J.~E.~Palomar and E.~Oset,
arXiv:hep-ph/0306188.

\bibitem{WuBES}
N.~Wu,
arXiv:hep-ex/0104050.

\bibitem{Augustin}
J.~E.~Augustin {\it et al.}  [DM2 Collaboration],
Nucl.\ Phys.\ B {\bf 320} (1989) 1.

\bibitem{Falvard}
A.~Falvard {\it et al.}  [DM2 Collaboration],
Phys.\ Rev.\ D {\bf 38} (1988) 2706.

\bibitem{Lockman}
W.~S.~Lockman  [MARK-III Collaboration],
SLAC-PUB-5139
{\it Presented at 3rd Int. Conf. on Hadron Spectroscopy, Ajaccio, France, Sep 23-27, 1989}


\end{thebibliography}
\end{document}